\documentclass[twoside,preprintnumbers,amsmath,amssymb,pacs,twocolumn,showpacs,nofootinbib]{revtex4}
\usepackage{amsmath,amssymb,url}
\usepackage{graphicx,pslatex,subfigure}

\usepackage{braket}
\usepackage{fancyvrb}
\newcommand{\omu}{\ensuremath{\overline \mu}}

\renewcommand{\d}{\ensuremath{\mathrm{d}}}

\renewcommand{\d}{\ensuremath{\mathrm{d}}}

\newcommand{\p}{\partial}

\newcommand{\e}{\ensuremath{\mathrm{e}}}

\begin{document}

\title{{\bf K\"all\'{e}n-Lehmann spectroscopy for (un)physical degrees of freedom}}
\author{David Dudal$^\dag$, Orlando Oliveira $^\ddag$, Paulo J.~Silva$^\ddag$}
\email{david.dudal@ugent.be, orlando@fis.uc.pt, psilva@teor.fis.uc.pt}
\affiliation{$^\dag$ Ghent University, Department of Physics and Astronomy, Krijgslaan 281-S9, 9000 Gent, Belgium\\ $^\ddag$ Centro de F\'{i}sica Computacional, Departamento de F\'{i}sica, Universidade de Coimbra, 3004-516 Coimbra, Portugal}

%\date{\today}
\begin{abstract}
We consider the problem of ``measuring'' the K\"all\'en-Lehmann spectral density of a particle (be it elementary or bound state) propagator by means of $4d$ lattice data. As the latter are obtained from operations at (Euclidean momentum squared) $p^2\geq 0$, we are facing the generically ill-posed problem of converting a limited data set over the positive real axis to an integral representation, extending over the whole complex $p^2$-plane. We employ a linear regularization strategy, commonly known as the Tikhonov method with Morozov discrepancy principle, with suitable adaptations to realistic data, e.g.~with unknown threshold. An important virtue over the (standard) maximum entropy method is the possibility to also probe unphysical spectral densities, as, for example, of a confined gluon.  We apply our proposal here to ``physical'' mock spectral data as a litmus test and then to the lattice $SU(3)$ Landau gauge gluon at zero temperature.
\end{abstract}
\maketitle
In most practical $4d$ quantum field theory computational schemes to date, be it continuum Feynman-diagrammatic (not necessarily perturbation theory) or lattice Monte Carlo based, an Euclidean setting is used. An obvious drawback is that physics happens in Minkowski space, so an analytic continuation is in order.  To name only a few examples where such effort is needed: (i) transport properties, which in general describe the response to a small external disturbance that drives the system a little bit out of its equilibrium state, making it an inherently (time) dynamical problem. (ii) particle properties, e.g.~a mass as a pole of a propagator does not show up in Euclidean correlators for $p^2\geq 0$ but for $p^2\leq 0$. This becomes particularly relevant for the bound state equations of particles that in se are not physically observable, e.g.~confined colored quarks and gluons \cite{Alkofer:2000wg,GimenoSegovia:2008sx}.

The problem we are thus facing is the analytic continuation of a function that is only known over the positive real axis, or even more only known in a limited set of data points on that semi-axis, for example obtained from a lattice QCD computation or a numerical solution to the quantum Dyson-Schwinger equations (DSE) of motion \cite{Alkofer:2000wg,Boucaud:2011ug}. One can always match a polynomial to the data set, but nobody will proclaim that all observable physics is embedded in a polynomial description. In the absence of some a priori global information as e.g.~location of cuts, the numerical analytical continuation is clearly an extremely ill-defined problem. Attempts have been made, as in \cite{GimenoSegovia:2008sx}, using the local Cauchy-Riemann equations obeyed by an analytic function, but the numerical stability remained a problem. Luckily, in several cases the indispensable a priori information is available. Let us consider $\mathcal{G}(p^2)\equiv\braket{\mathcal{O}(p)\mathcal{O}(-p)}$, the Euclidean momentum-space propagator of a (scalar) physical degree of freedom\footnote{Straightforward generalizations apply to states with spin. Integral representations for finite temperature correlation functions can be found in e.g.~\cite{Meyer:2011gj}.}, then it must have a K\"{a}ll\'{e}n-Lehmann (KL) spectral representation form, see e.g.~\cite{Peskin:1995ev},
\begin{equation}\label{KL1}
    \mathcal{G}(p^2)=\int_{0}^{+\infty}\d\mu\frac{\rho(\mu)}{p^2+\mu}\,,\qquad \text{with }\rho(\mu)\geq0 \text{ for } \mu\geq 0\,.
\end{equation}
The spectral density,
\begin{equation}\label{2}
\rho(\mu)=\sum_\ell \delta(\mu- m_\ell^2) \left|\left\langle0|\mathcal{O}|\ell_0\right\rangle\right|^2\,,
\end{equation}
contains information on the masses of physical states described by the operator  $\mathcal{O}$ (isolated $\delta$-function contributions), as well as on where the multiparticle spectrum sets in, defining a threshold. Here $\left|\ell_0\right\rangle$ refers to all states\footnote{The $\sum_\ell$ is a symbolic notation as there is a continuum of states, next to a discrete (possibly bound state) spectrum.} at rest ($\vec{p}=0$) that have an overlap with the operator $\mathcal{O}$. Obviously, $\rho(\mu)\geq0$. Moreover, \eqref{KL1} defines a function over the complex Euclidean $p^2$ plane that is everywhere analytic except for a branch cut for real $p^2\leq 0$. From the KL representation, we learn $\rho(\mu)\propto \text{Disc}_{\mu\geq0} \text{Im}[\mathcal{G}(-\mu)]$, while from the optical theorem \cite{Peskin:1995ev}, $\text{Im}[\mathcal{G}(\mu)]\propto \text{cross section}$, giving a clear physical reason behind $\rho(\mu)\geq 0$. Given the prescribed integral form \eqref{KL1}, the problem is reduced to finding $\rho(\mu)$ given data input for $\mathcal{G}(p^2)_{p^2\geq 0}$. This is still an ill-posed problem, best appreciated when considered in terms of the inverse Laplace transform. Indeed, with $\mathcal{L}$ the Laplace integral, $F(t)=(\mathcal{L}f)(t)\equiv\int_0^{+\infty}\d s e^{-st}f(s)$, eq.~\eqref{KL1} can be reexpressed as
\begin{eqnarray}\label{sub2}
  \mathcal{G}=\mathcal{L}^2\hat\rho=\mathcal{L}\mathcal{L}^\ast \hat\rho\,.
\end{eqnarray}
We introduced the adjoint $\mathcal{L}^\ast$ of the Laplace-operator, being $\mathcal{L}$ itself. As taking $\mathcal{L}^{-1}$ is a notorious ill-posed problem due to the exponential dampening, quite obviously so is the double inversion.  For $\mathcal{G}(p^2)$, we have usually a set of data points with error bars. Let us assume that
\begin{equation}\label{sub3}
    ||\mathcal{G}-\mathcal{G}_\delta||\leq \delta\,,
\end{equation}
where $||\cdot||$ represents either the continuum $L_2$ norm, $||f||=\sqrt{\int |f|^2}$, or the usual Euclidean vector norm in a discrete setting. For the applications discussed we define $\delta=\sqrt{\sum{(\text{errors})^2}}$. Thus, eq.~\eqref{sub3} expresses that we have unprecise data for $\mathcal{G}(p^2)$ within a ``noise'' level $\delta$. The essence of an ill-posed problem is that very small variations in the input data for $\mathcal{G}$ can cause utterly violent changes in the output. Overcoming it boils down to finding an estimate for $\rho$ such that when $\delta\to0$, the corresponding approximate solution goes to the exact $\rho$. In order to handle this, one usually looks at a properly \emph{regularized} version of the problem. Such problems are well-studied in signal processing sciences and in the context of spectral representations of Green functions (at finite temperature), it has become popular to rely on the maximum entropy method (MEM) \cite{Asakawa:2000tr}.

We will now develop an alternative to MEM for inverting the KL representation. Preliminary attempts can be found in \cite{Oliveira:2012eu}. We first provide a few general concepts, loosely following \cite{inverse} whereto we refer for the mathematical background. Consider a generic ill-posed problem with operator $\mathcal{K}$:
\begin{equation}\label{tik1}
    y=\mathcal{K}x\,,\quad ||y-y_\delta||\leq \delta\,.
\end{equation}
Standard Tikhonov regularization amounts to search for the solution $x_\lambda$ wherefore
\begin{equation}\label{tik2}
    \mathcal{J}_\lambda=||\mathcal{K}x-y||^2+\lambda||x||^2
\end{equation}
is minimal; $\lambda>0$ is a regularization parameter. Notice that in the above the 2 norms can be differently chosen. $x_\lambda$ is obtained as the solution of the so-called normal equation \cite{inverse}
\begin{equation}\label{tik3}
    \lambda x_\lambda +\mathcal{K}^\ast \mathcal{K} x_\lambda =\mathcal{K}^\ast y\,.
\end{equation}
The operator $\lambda+\mathcal{K}^\ast \mathcal{K}$ is strictly positive, hence invertible, thus \eqref{tik3} ought to have a unique solution. The ill-posedness of the original problem, \eqref{tik1}, is cured for $\lambda>0$. Indeed, ill-posed problems can be traced back to having near-to-zero singular values of $\mathcal{K}$, when a singular value decomposition is employed. The regularization parameter screens the too small singular values and lies at the very heart of obtaining a well-defined problem\footnote{MEM is  actually a special case of a more general Tikhonov regularization strategy whereby $||x||^2$ is replaced by a--- not necessarily quadratic in $x$--- penalty function $\Psi(x)$. MEM corresponds to using e.g.~$\psi=-\int x\ln x$, the Shannon-Jaynes entropy. The nonlinear nature of this regularization, given the presence of the $\ln$, makes it computationally burdensome. In our current alternative approach, we aim at a simpler (and thus more computation-friendly) regularization.}. To fix $\lambda$, we will resort to an a posteriori fixing, by making use of the solution $x_\lambda$: the Morozov discrepancy principle \cite{inverse}; one chooses that particular $\overline\lambda$ which gives
\begin{equation}\label{tik4}
    ||\mathcal{K}x_{\overline\lambda}-y_\delta||=\delta\,.
\end{equation}
A unique solution $x_{\overline\lambda}$ to \eqref{tik4} exists \cite{inverse}. This particular choice is reasonable: if the noise on the input data vanishes, $\delta\to0$, the ``noise'' on the approximate equation will also vanish. In a sense, \eqref{tik4} expresses that we aim for ``output'' of similar quality as the ``input''. Simultaneously, the discrepancy principle avoids selecting a too small $\lambda$, which would drive us back to the ill-posed case. In the continuum, the solution converges to the exact one for vanishing noise \cite{inverse}.

We will now adapt to a discrete setting the inversion of the integral equation
\begin{equation}\label{KL1b}
    \mathcal{G}(p^2)=\int_{\mu_0}^{+\infty}\d\mu\frac{\rho(\mu)}{p^2+\mu}\,,
\end{equation}
similar as was done in \cite{thesis2} for the Laplace transform. Notice that we introduced a to be determined threshold $\mu_0$ into the integral definition. Setting $\mathcal{G}_i\equiv\mathcal{G}(p_i^2)$ and assuming we have $N$ data points, we need to minimize
\begin{equation}\label{tikdis1}
  \mathcal{J}_\lambda=\sum_{i=1}^{N}\left[\int_{\mu_0}^{+\infty} \d\mu\frac{\rho(\mu)}{p_i^2+\mu}-\mathcal{G}_i\right]^2+\lambda \int_{\mu_0}^{+\infty} \d\mu~\rho^2(\mu)
\end{equation}
where we took into account that the KL integral operator is self-adjoint. Perturbing $\rho(\mu)$ linearly and demanding that the variation of $\mathcal{J}_\lambda$ vanishes, leads to the normal equation
\begin{equation}\label{tikdis2}
  \sum_{i=1}^{N} \underbrace{\left[\int_{\mu_0}^{+\infty}\d\nu\frac{\rho(\nu)}{p_i^2+\nu}-\mathcal{G}_i\right]}_{\equiv c_i}\frac{1}{p_i^2+\mu}+\lambda\rho(\mu)=0\,\, (\mu\geq\mu_0)
\end{equation}
after some rearranging. Said otherwise, the (regularized) solution to KL inversion is explicitly given by
\begin{equation}\label{tikdis3}
  \rho_{\lambda}(\mu)=-\frac{1}{\lambda}\sum_{i=1}^{N}\frac{c_i}{p_i^2+\mu}\theta(\mu-\mu_0)\,,
\end{equation}
with $\theta(\cdot)$ being the step function. Notice that the threshold is crucial to avoid a singularity at $\mu=0$ if $\mathcal{G}(p_i^2=0)<\infty$ would be part of the inversion. The $c_i$ are evidently still in order to give meaning to the foregoing equation. Combination of eqns.~\eqref{tikdis2},\eqref{tikdis3} yields
\begin{equation}\label{tikdis4}
  c_i=-\frac{1}{\lambda}\int_{\mu_0}^{+\infty}\d\nu\frac{1}{p_i^2+\nu}\sum_{j=1}^{N}\frac{1}{p_j^2+\nu}c_j-\mathcal{G}_i\,,
\end{equation}
i.e.~a linear system of equations
\begin{equation}\label{tikdis5}
 \lambda^{-1} \mathcal{M}c+c=-\mathcal{G}\,,
\end{equation}
with
\begin{equation}\label{tikdis6}
  \mathcal{M}_{ij}=\int_{\mu_0}^{+\infty}\d\nu\frac{1}{p_i^2+\nu}\frac{1}{p_j^2+\nu}=\frac{\ln\frac{p_j^2+\mu_0}{p_i^2+\mu_0}}{p_j^2-p_i^2}\,.
\end{equation}
As $\mathcal{M}_{ii}=1/(p_i^2+\mu_0)$, we have a perfectly well-defined, symmetric matrix for $\mu_0>0$. The inverse KL operation has been reduced to solving a linear system of equations, \eqref{tikdis5}, in terms of which the solution the spectral density is given by eq.~\eqref{tikdis3}. Moreover, to implement the Morozov discrepancy principle \eqref{tik4}, we notice that the reconstructed propagator can be directly expressed in terms of the $c_i$:
\begin{equation}\label{tikdis7}
  \mathcal{G}_{\lambda}(p^2)=\int_{\mu_0}^{+\infty}\d\mu\frac{\rho_{\lambda}(\mu)}{p^2+\mu}=-\frac{1}{\lambda}\sum_{i=1}^{N}\frac{c_i\ln\frac{p^2+\mu_0}{p_i^2+\mu_0}}{p^2-p_i^2}\,.
\end{equation}
Since the threshold $\mu_0$ is a priori free, we will use the optimal (Morozov) regulator $\overline\lambda$, which depends on $\mu_0$, to fix it: we will look for a region of stability (i.e.~minimum) in $\overline\lambda(\mu_0)$. This is a natural criterion, since the smaller $\lambda$ the closer we are to the original problem.

An important remark is still in order. The formal solution \eqref{tikdis3} implicitly assumes that $\rho(\mu)\sim 1/\mu$ for $\mu$ large. This is however not always the case, depending on the correlator $\mathcal{G}(p^2)$ that is being investigated. In asymptotically free gauge theories, (RG improved) perturbation theory can be assumed valid at large momenta, under which conditions the spectral density, for large values of its argument, can be estimated directly via $\rho(\mu)\propto \text{Disc}_{\mu\geq0} \text{Im}[\mathcal{G}(-\mu)]$. This large $\mu$-behaviour can then be superimposed onto the above analysis by adding a weight to the last integral appearing in eq.~\eqref{tikdis1}, corresponding to choosing an appropriate norm in eq.~\eqref{tik2} for $||x||^2$, i.e.~on the space of suitably tempered spectral functions. Most of the foregoing computations carry over, the inverse weight enters the solution \eqref{tikdis3} thereby producing the desired asymptotic behaviour. We plan to come back to this in a more extensive work.

As a first application, let us consider a (non-relativistic) ``Breit-Wigner'' toy spectral density with nonzero threshold,
\begin{eqnarray}\label{bw1}
  \rho(\mu)&=&\frac{\mu}{(\mu-m^2)^2+\Gamma^2/4}\theta(\mu-\omu)\\
  && m^2=1~\text{GeV}^2\,, \Gamma=1~\text{GeV}^2\,,\omu=0.1~\text{GeV}^2\,.\nonumber
\end{eqnarray}
The propagator corresponding to eq.~\eqref{bw1} was computed using a Gauss-Legendre quadrature with 1000 points and $\sqrt{\mu}_{max}=20$~GeV. We checked robustness against a change of the number of Gauss-Legendre points. The system \eqref{tikdis5} was solved using a Gauss-Jordan normal elimination with $N=120$ entry data points. We assigned to each data point $\mathcal{G}_i$ the percent errors $\varepsilon=10, 5, 1, 0.1, 0.001, 0.0001$ according to $\mathcal{G}_i\times\varepsilon\times(0.5+0.5r)$, with $r$ a uniform random number $\in[0,1]$. The propagator ($=$ mock data) and its reconstructions from the spectral functions are shown in FIG.~(1a), resp.~(1b).
\begin{figure}[h]
  \begin{center}
   \subfigure[Input propagators and reconstructions.]{\includegraphics[width=4.1cm]{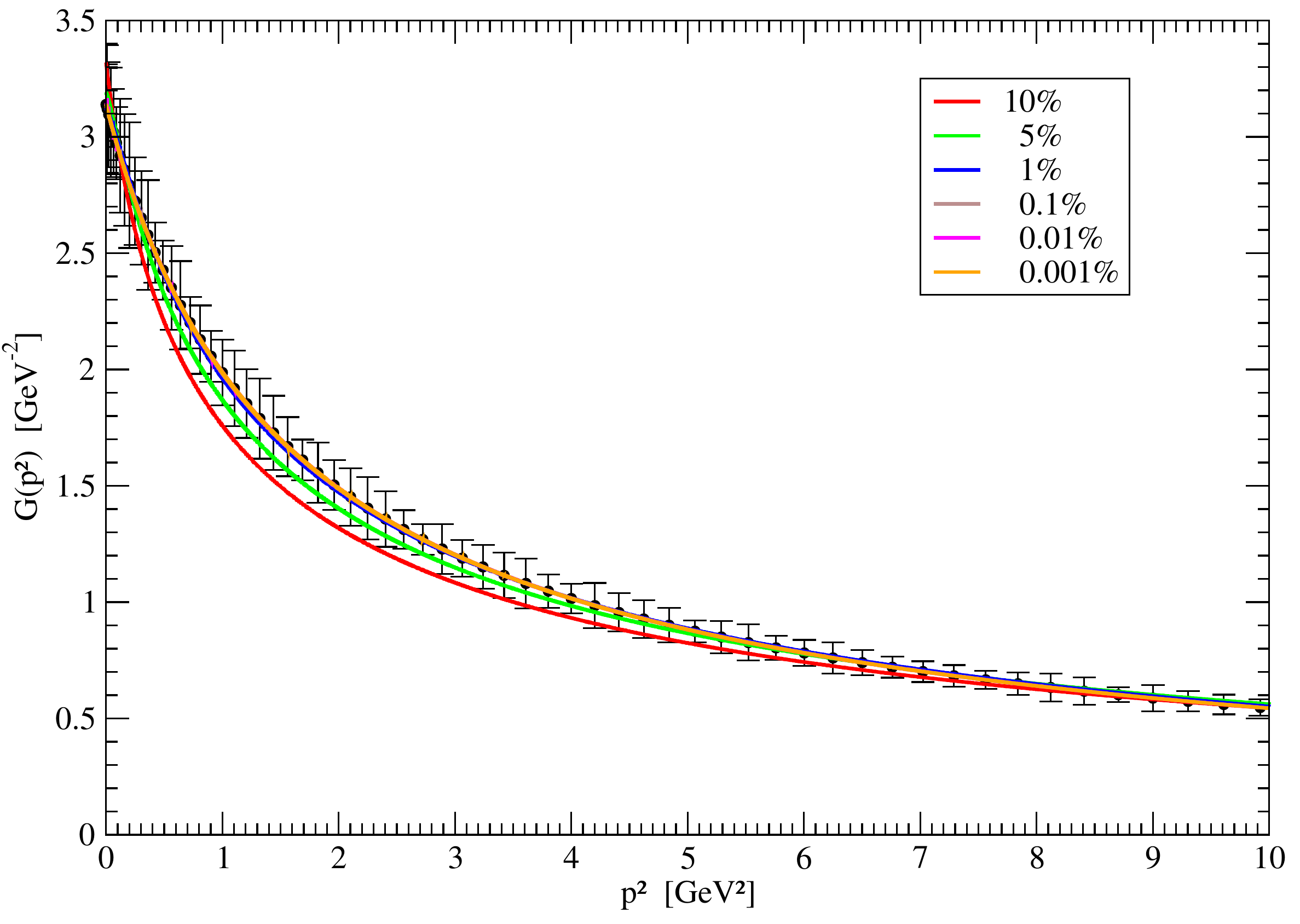}\label{fig1a}}
%      \hspace{-0.4cm}
    \subfigure[Spectral function and $\rho(\mu)$ from the inversion in terms of errors.]{\includegraphics[width=4.1cm]{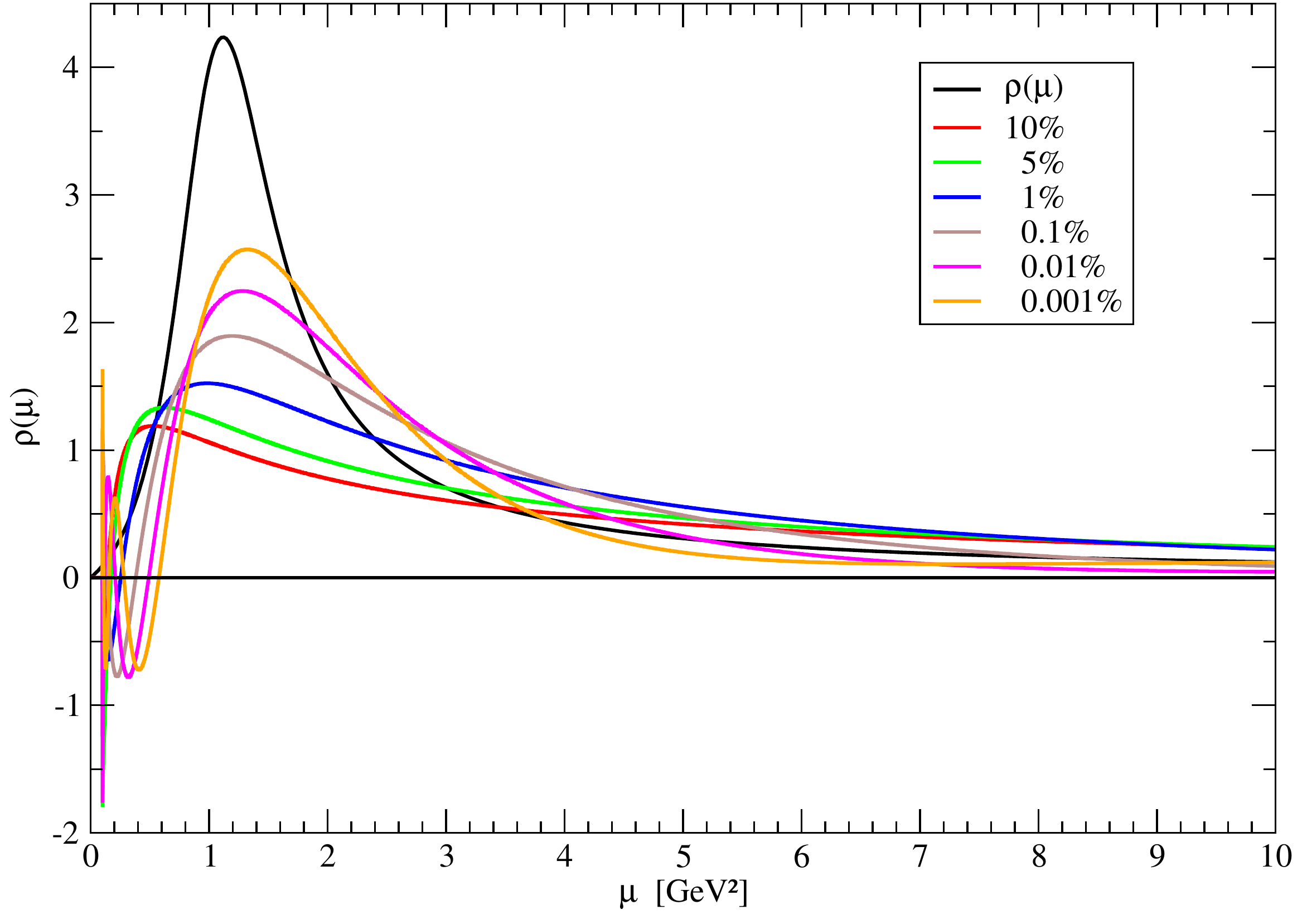} \label{fig1b}}
  \end{center}
 \caption{The Breit-Wigner toy model with optimal $\mu_0=0.1~\text{GeV}^2$.}
\end{figure}
For the optimal threshold, we refer to FIG.~(2a) where, quite surprisingly, we find that $\mu_0\approx 0.1~\text{GeV}^2$, i.e.~at the location of the exact threshold $\omu$. This gives credit to our criterion. We opted for a 1\% error margin in the input data here to mimic a somewhat realistic situation.
\begin{figure}[h]
  \begin{center}
   \subfigure[Breit-Wigner.]{\includegraphics[width=4.1cm]{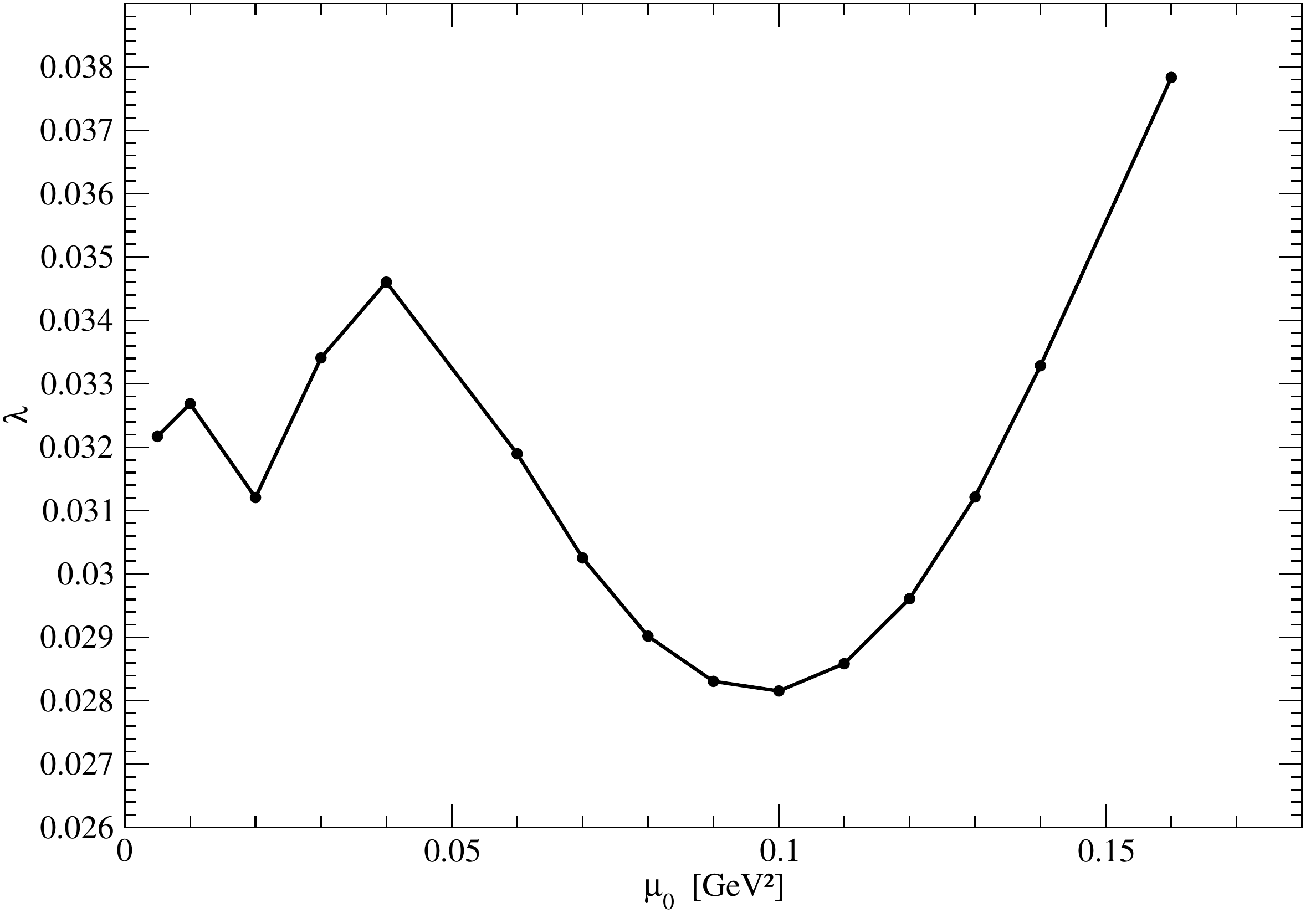}\label{fig2a}}
%      \hspace{-0.4cm}
    \subfigure[Gluon.]{\includegraphics[width=4.1cm]{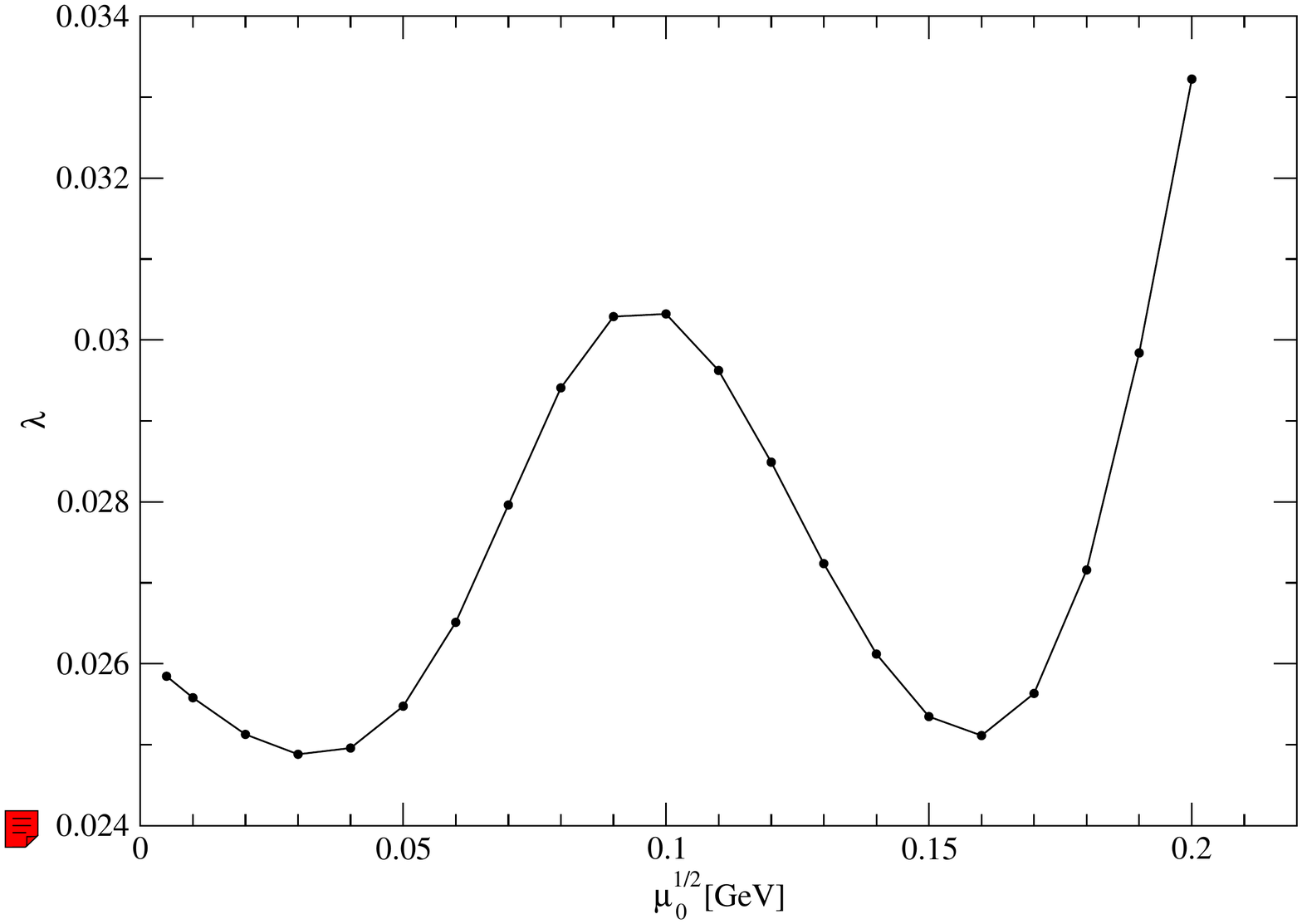} \label{fig2b}}
  \end{center}
 \caption{The Morozov parameter $\overline\lambda$ in terms of $\mu_0$.}
\end{figure}
From Figs.~(1a),(1b) we observe that the inversion method is capable of reproducing a peak in the right area, with increasing height if the noise on the input data gets smaller ($\delta\to0$). The quality of the reproduced propagator starts to be excellent for errors of the order of 1\% or smaller. It is worthwile noticing that these features of our inversion method are very similar to those found with a MEM analysis, see in particular \cite[FIG.~4]{Asakawa:2000tr}.

We have also investigated the effect of choosing a $\mu_0$ slightly different from the optimal one. We can report that the main difference appeared in the deep infrared region (very small $p^2$) of the reconstructed propagator.

Next, we turn to a real example: we consider contemporary lattice data for the pure (no quarks) $SU(3)$ Yang-Mills gluon, quantized in the Landau gauge $\p_\mu A_\mu=0$. The ensuing propagator can be written as $\mathcal{D}_{\mu\nu}(p^2)=\left(\delta_{\mu\nu}-\frac{p_\mu p_\nu}{p^2}\right)\mathcal{D}(p^2)$ due to its transverse nature. The data, discussed in e.g.~\cite{Oliveira:2012eh}, was obtained simulating the Wilson action for pure Yang-Mills theory at a $\beta = 6.0$, i.e. with a lattice spacing $a = 0.1016(25)$~fm, on a $80^4$ hypercubic lattice which has a physical volume of
$(8.13$ fm$)^4$. As it is well-known, this propagator displays a violation of positivity \cite{Cucchieri:2004mf}, made clear by probing the Schwinger function, defined via
\begin{equation}\label{KL4}
\Delta(t)\equiv\frac{1}{2\pi}\int_{-\infty}^{+\infty}\d p\,\e^{-i pt}\mathcal{D}(p^2)\left(=\int_{0}^{+\infty}\d y\,\rho(y^2)\e^{-ty}\right)\,;
\end{equation}
the bracketed eq.~assumes a KL representation for the gluon. As the gluon is not an observable asymptotic particle state, there is absolutely no guarantee it must display a KL representation. But if so, if $\Delta(t)$ is not positive, then neither can $\rho(t)$ be, showing that the gluon cannot be physical. This observation has been frequently used as a practical way to establish gluon confinement. Notice that if $\Delta(t)$ would be positive, we would not know anything on the sign of $\rho(t)$. As already pointed out before, if unphysical gluons are to be combined into physical bound states (viz.~the experimentally elusive glueballs \cite{Ochs:2013gi}), information on the precise analytic properties of the gluon propagator are desirable, see also \cite{Dudal:2010cd}. In a recent Letter \cite{Strauss:2012dg}, the gluon DSE was numerically solved in the complex plane using part of the machinery developed in \cite{Maris:1995ns}. A discontinuity along the negative real axis was observed, the corresponding jump can then be identified with the spectral function $\rho(\mu)$. The latter showed a quite peculiar form, starting positive but changing sign rapidly over a small momentum window with sharp but finite peak around $\sqrt{\mu}\sim 0.6~\text{GeV}$. A similar form, albeit with a $\delta$-function (thus infinite peak) where the function becomes negative was found in \cite{Iritani:2009mp} by \emph{fitting} $\mathcal{D}(p^2)_{p^2\geq0}\sim (p^2+m^2)^{-3/2}$. As solving the gluon DSE is a complicated task even for $p^2\geq0$ due to the infinite tower of equations, making truncations and modeling of the ingoing vertex interactions are necessary. It would be beneficial to use Euclidean lattice data to complement the DSE analysis. Our proposed methodology serves this goal exactly. Notice that (standard) MEM is out of the question here\footnote{There exist an extension to nonpositive spectral functions, albeit that the concept of ``entropy'' is actually lost in such cases \cite{generalizedMEM}.}, since that relies on the a priori positivity of $\rho(\mu)$ and its use as a probability function \cite{Asakawa:2000tr}. The same comment applies to a recent alternative to MEM \cite{Burnier:2013nla}. A simple dimensional analysis learns that  $\rho(\mu)\sim_{\mu\to\infty} 1/\mu$, consistent with \eqref{tikdis3}. The (computable) logarithmic corrections to this estimate can be taken into account in a more complete analysis later on.
\begin{figure}[t]
  \begin{center}
\includegraphics[width=6cm]{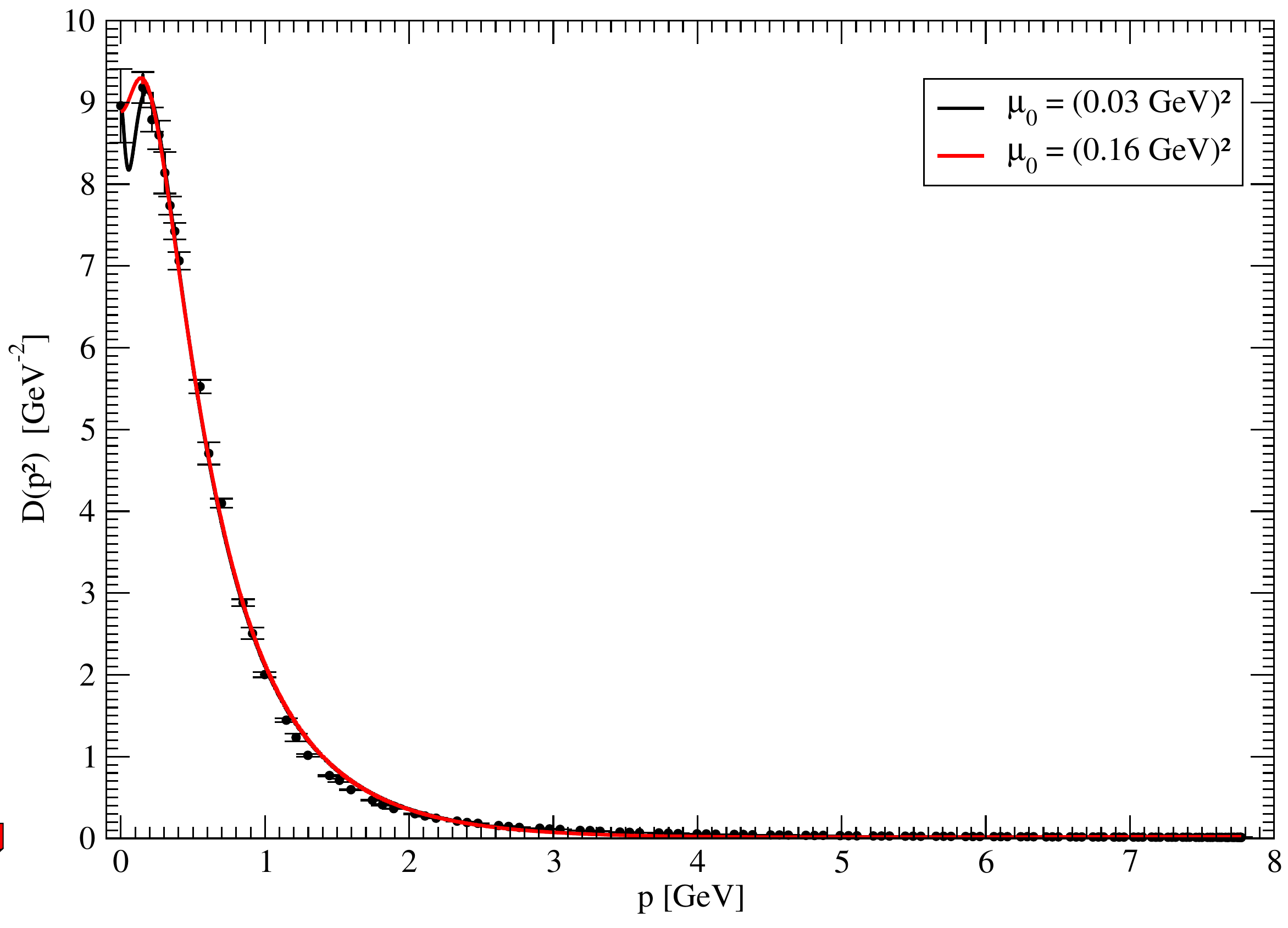}\label{fig3}
\end{center}
 \caption{Gluon propagator and reconstruction.}
\end{figure}
We renormalized the gluon lattice data in a MOM scheme at $\mu=4$ GeV for definiteness -- see~\cite{Oliveira:2012eh} for details.
The highest momenta accessed by our simulation is $p_{max}=7.77~\text{GeV}$. The number of lattice data points is $124$ and the noise level is set by $\delta=0.658~\text{GeV}^{-2}$. In FIG.~(2b) we notice the occurrence of 2 minima for $\overline\lambda(\mu_0)$, at $\mu_0\approx0.03~\text{GeV}$ and $\mu_0\approx0.16~\text{GeV}$, with the former one giving a slightly lower value of $\overline\lambda$. For both values, the reconstructed propagator and associated spectral density are shown in Figs.~(3), (4a) and (4b). The differences in the spectral density translate mostly into a different deep IR behaviour of the reconstructed gluon propagator. The main observation however is that the gluon spectral density is indeed a nonpositive quantity. One can also compare our estimate for the gluon spectral function, based on lattice data, with the numerical output of solving the complex momentum DSE, \cite[FIG.~5]{Strauss:2012dg}. With our current results, we do not see evidence of the reported sharp peak, while the violation of positivity sets in already for small $\mu$, rather than after $\mu\sim(0.6)^2~\text{GeV}^2$ \cite{Strauss:2012dg}.

\begin{figure}[h]
  \begin{center}
   \subfigure[Linear scale.]{\includegraphics[width=4.1cm]{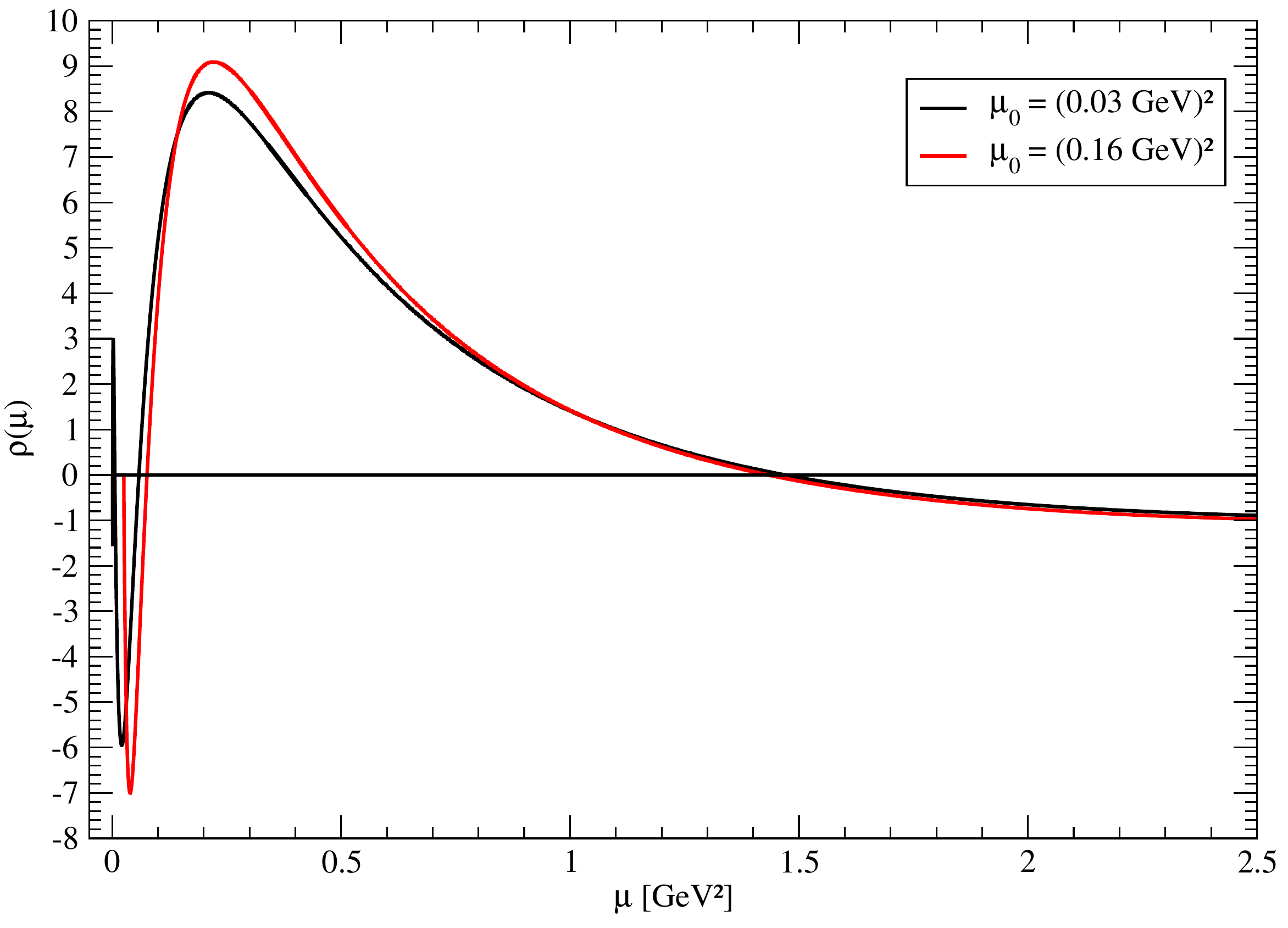}\label{fig4a}}
%      \hspace{-0.2cm}
    \subfigure[Log scale.]{\includegraphics[width=4.1cm]{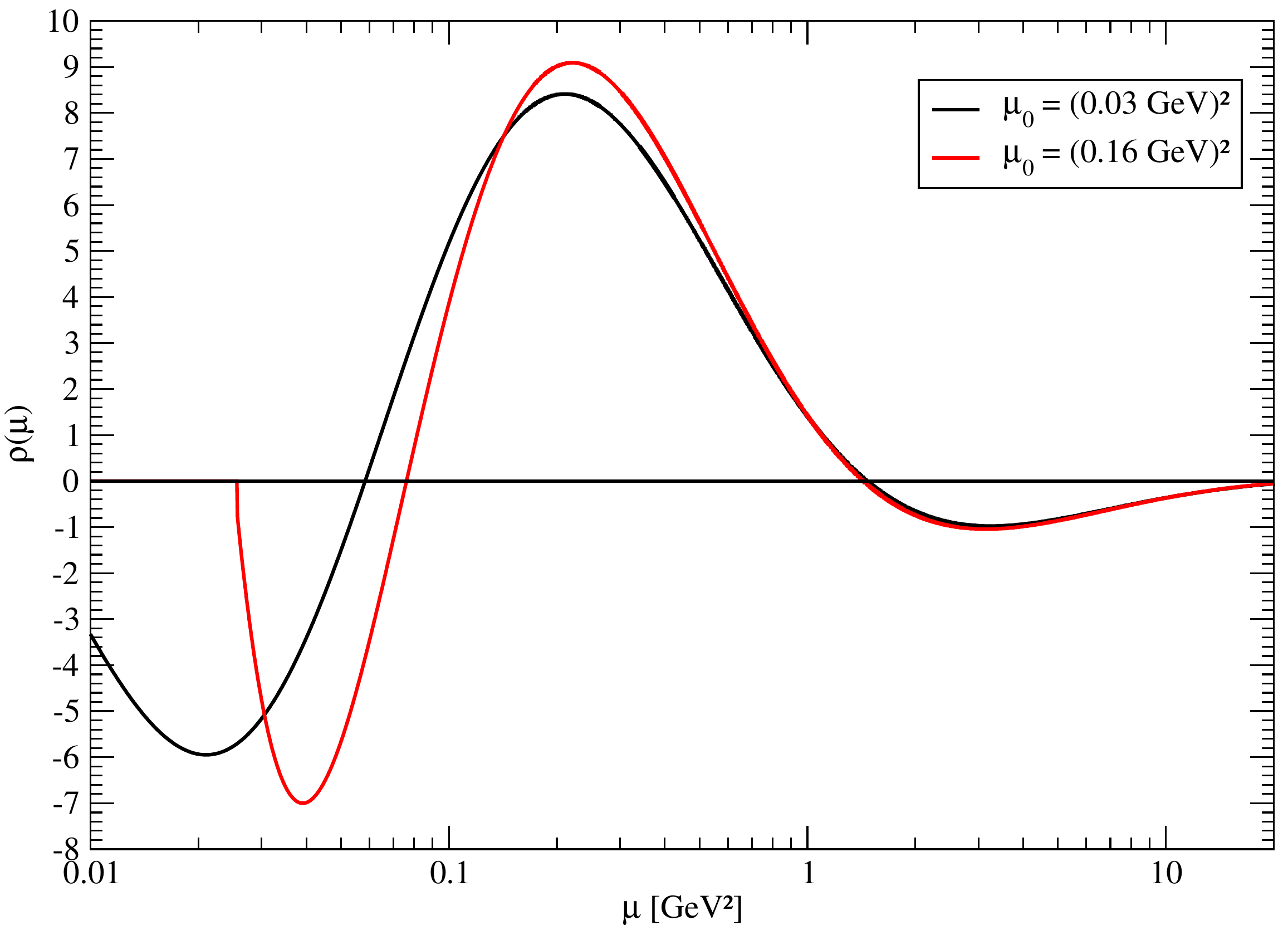} \label{fig4b}}
  \end{center}
 \caption{Gluon spectral function.}
\end{figure}
In conclusion, we have presented a linear regularization strategy to numerically probe spectral densities of two-point correlation functions. In work in progress, we are testing the method on the physical $SU(3)$ lattice scalar glueball, which outcome can be tested against independently obtained mass estimates \cite{Morningstar:1999rf,Oliveira:2012eu}. In the unphysical glue sector, we are also studying into more depth the gluon propagator no longer assuming a cut along the negative real axis, but rather using rational (Pad\'{e}) approximation theory and the phenomenon of Froissart doublets \cite{biomet} to get possible insights into where the branch points could be located, whereafter by suitably deforming the branch cut a spectral analysis with the tools from this paper would become feasible. We foresee future applications in the quark sector and finite temperature QCD, to study e.g.~the spectral properties of the electric and magnetic gluons.

\emph{D.D.~is partially supported by the Research Foundation-Flanders via the Odysseus grant of F.~Verstraete; P.J.S.~by FCT under contract SFRH/BPD/40998/2007 and O.O., P.J.S.~by the FCT research projects
CERN/FP/123612/2011 and  PTDC/FIS/100968/2008, developed under the initiative QREN financed by the UE/FEDER through the Programme COMPETE - Programa Operacional
Factores de Competitividade.}

\end{document}